# Material model calibration through indentation test and stochastic inverse analysis

Vladimir Buljak[1], Shwetank Pandey[1]

[1]*Department of strength of materials – Faculty of mechanical engineering, University of Belgrade, Belgrade, Serbia*

ABSTRACT: Indentation test is used with growing popularity for the characterization of various materials on different scales. Developed methods are combining the test with computer simulation and inverse analyses to assess material parameters entering into constitutive models. The outputs of such procedures are expressed as evaluation of sought parameters in deterministic sense, while for engineering practice it is desirable to assess also the uncertainty which affects the final estimates resulting from various sources of errors within the identification procedure. In this paper an experimental-numerical method is presented centered on inverse analysis build upon data collected from the indentation test in the form of force-penetration relationship (so-called 'indentation curve'). Recursive simulations are made computationally economical by an 'a priori' model reduction procedure. Resulting inverse problem is solved in a stochastic context using Monte Carlo simulations and non-sequential Extended Kalman filter. Obtained results are presented comparatively as for accuracy and computational efficiency.

## 1. Introduction

Indentation tests are employed since decades for mechanical characterization of materials at diverse scales and in various technological fields. The original form of the test was fairly simple, intended for the measurement of 'hardness' in metals. Later the assessment of mechanical parameters entering into constitutive models and necessary for overall inelastic structural analyses became the purpose of this test. The parameter identification at present is usually based on 'instrumented' indenters, namely those that are capable to provide a so-called 'indentation curve', a curve that correlates the indenter tip penetration versus force applied on it, see Figure 1.

Transition from indentation curves to the sought parameters in the early works was accomplished by semi-empirical formulae (see e.g. [1]). Recently the use of computational methods through 'inverse analysis' has been gaining popularity and is a subject of vast literature (see e.g. [2, 3, 4]). In this concept, the test is simulated, traditionally by the use of Finite Element Modeling (FEM), while the results of simulation are compared to those measured in the experiment in the form of a 'discrepancy function'. This function quantifies the difference between computed and measured data, and in a subsequent phase it is minimized with respect to sought parameters by a suitable mathematical programming technique.

Material parameter calibration based on inverse analysis offers improved accuracy and better flexibility with respect to semi-empirical formulae. This is particularly the case when uncertain or incomplete data, are used from the experiment. In practical applications of material model calibration based on indentation test, this circumstance occurs quite frequently since there can be many factors affecting the measured indentation curve (e.g. rough surface of tested specimen, error in measured force and corresponding displacement, slight difference in local material properties etc, [5]).

In order to use measurable quantities endowed by a covariance matrix quantifying their uncertainties, it is required to design an inverse analysis procedure capable to estimate the propagation of this uncertainty up to the assessed parameters. To achieve such goal, a stochastic framework needs to be adopted for the resulting inverse problem.

In this paper Monte Carlo (MC) simulations and non-linear Kalman filters (KF) are employed to solve the parameter identification problem in a stochastic context, on the basis of indentation curves generated by an instrumented indenter. The indentation test referred to in this study considers conical indenter with $120^0$ cone opening angle and a 200μm spherical radius on a tip, like in traditional Rockwell tests [6]. The material specimen response is assumed to be isotropic elastic-plastic with linear hardening, following Drucker-Prager yield criterion in a view of potential application of proposed approach to the characterization of metal-ceramic composites.

The developed procedure is verified within pseudo-experimental context (i.e. computer generated data are used instead of those collected from a real experiment) in order to comparatively check different strategies in terms of accuracy and computational efficiency in the present context. Computational burden connected to recursive simulations of an indentation test, particularly required by Monte Carlo simulations, is significantly reduced by the employment of a "model reduction" technique based on Proper Orthogonal Decomposition (POD) apt for the acceleration of non-linear simulations (see [7] for details).

The paper is organized as follows: Section 2 gives a brief outline of inverse analysis by mathematical programming and model reduction technique for fast and economical parameter identification. Section 3 is devoted to the stochastic approaches described in the form implemented in present context. A reference problem with several specific modeling features adopted in this study is given in Section 4, followed by comparative presentation of achieved numerical results. Conclusions and future research prospects considering discussed, and related items are outlined in Section 5.

## 2. Inverse analysis with mathematical programming and model reduction

Material model calibration based on inverse analysis relies on performing of an experiment (suitably chosen to provide measurable quantities sensitive to the sought parameters) and its subsequent simulation. Results gathered from the experiment are used together with their computed counterpart in order to form a 'discrepancy function' that quantifies the difference between the two. Resulting inverse analysis is centered on minimization of this function. Such problem can be briefly formulated as follows.

Let $\mathbf{u}_e$ be the vector of experimentally measured data, and $\mathbf{u}(\mathbf{p})$ the computed counterpart resulting from a simulation, with sought parameters collected in vector $\mathbf{p}$. The discrepancy function in a purely deterministic formulation in its simplest form is given by:

$$\omega(\mathbf{p}) = \left[\mathbf{u}_e - \mathbf{u}(\mathbf{p})\right]^T \cdot \left[\mathbf{u}_e - \mathbf{u}(\mathbf{p})\right] \qquad (1)$$

The solution to the formulated inverse problem is defined by a set of parameters collected in vector $\mathbf{p}_{SOL}$ that is minimizing the above defined discrepancy function. This minimization problem is solved numerically as there is no analytical description of the objective function (1). To this purpose one can make a recourse to 'first order' algorithms which are relying on a quadratic approximation of the objective function. Example of these are methods based on 'Trust Region Algorithm' (TRA). Description of various TRA types is widely available in the literature (see e.g. [8, 9]) and a detailed presentation is omitted here for brevity. However, a brief outline of one of the most popular versions of TRA, used also in this study, is given in what follows for the sake of completeness.

In each iteration of TRA a quadratic approximation of the objective function is generated by an approximation of the Hessian matrix computed through Jacobian of the 'residual' vector, namely a vector that collects differences between measurable and computed quantities. Such quadratic approximation of the objective function represents a model function that is 'trusted' to be an adequate representation of the real function within defined region. Minimization of this model function is computationally inexpensive and its minimizer is searched for within a two-dimensional subspace spanned by gradient and by the Newton vector of the discrepancy function in the current iterate point defined by $\mathbf{p}_i$. The solution of this sub-problem minimization, denoted by $\mathbf{p}_{i+1}$, reduces the value of the objective function and provides a new point in which quadratic model function can be generated. This process is iteratively repeated until convergence criteria are met, and the procedure is terminated in the point which represents the global minimizer of the objective function. Repeated diverse initializations have fragmentarily to be performed in order to avoid ending of the step sequence in a local minimum due to possible lack of convexity.

Employing such algorithm requires calculation of first derivatives of the objective function with respect to each of the sought parameters. These derivatives are computed by finite differences, requiring in each iteration $P+1$ simulation of the test, where $P$ is the number of sought parameters. This circumstance makes the characterization procedure computationally lengthy, particularly in a presence of large non-linearities like in a simulation of indentation test.

Remarkable practical advantages can be achieved by invoking a model reduction technique based on Proper Orthogonal Decomposition in order to accelerate simulations. The POD procedure has been already adopted in researches and its description can be found in [10, 11]. Here just a brief outline of stages involved in it will be given.

Within 'search domain' in the sought parameter space $N$ points are selected that are uniformly covering this domain. Each one of this $N$ 'nodes' is assumed as input to the simulation of test leading to computed outputs collected in vector $\mathbf{u}_i$. Exploiting "correlation" of the $N$ calculated vectors a new basis is computed for them and its axes with negligible components of vectors $\mathbf{u}_i$ are dropped. Each test output $\mathbf{u}_i$ is now approximated by "amplitude" vector $\mathbf{a}_i$, that is its projection to sub-space with reduced dimensionality spanned by orthogonal basis denoted by $\bar{\mathbf{\Phi}}$. The orthogonal basis is generated by the eigenvalue computation of symmetric positive-semi-definite matrix of the order $N$. Starting from any new parameter vector $\mathbf{p}$, the corresponding vector $\mathbf{a}$ and the related computed response to a test $\mathbf{u}$, can now be calculated with controllable accuracy and with much smaller computational effort by means of Radial Basis Functions (RBF) interpolation among the previously computed responses $\mathbf{u}_i$ with parameters $\mathbf{p}_i$ ($i=1,2, \ldots , N$).

The above model reduction procedure makes the parameter identification fast and economical, as the recursive simulations required by the optimization algorithm are performed using 'light' computational tool.

Using numerical optimization algorithms like TRA the output of resulting inverse problem is a vector of parameters with their estimated values. In those cases when experimental data are endowed by corresponding covariance matrix, denoted by $\mathbf{C}^{EXP}$, quantifying the experimental error of measured quantities, the discrepancy function can be written in weighted residual form such as:

$$\omega(\mathbf{p}) = \left[\mathbf{u}_e - \mathbf{u}(\mathbf{p})\right]^T \cdot \mathbf{C}^{EXP} \cdot \left[\mathbf{u}_e - \mathbf{u}(\mathbf{p})\right] \qquad (2)$$

This modification is important in the situation where larger noise in measurements is present in certain zones (e.g. starting part of the indentation curve due to asperities of the unpolished surface). In this form of discrepancy function merely more importance is conferred to more accurate measurements, but the context remains deterministic, as the output of the minimization is still a single vector of assessed parameters without quantification of the uncertainty affecting them. In engineering applications however, it is highly desirable to estimate also the uncertainty of the identified parameters. To reach such result a stochastic procedure needs to be employed for solving the inverse problem. In the following section a brief outline of stochastic methods employed in this study is presented.

## 3. Stochastic inverse analysis for material parameter calibration

When uncertain measurable quantities, collected from the test, are used within inverse analysis procedure for material parameter calibration, it is important to quantify to which extent these uncertainties are affecting the outputs (namely sought parameters). To this purpose a stochastic procedure needs to be employed in order to solve the inverse problem. In this study two particular methodologies are used in order to reach such result: Monte Carlo sampling and Kalman filters. In what follows an outline of the application of these methods in the present context of material characterization based on indentation test is given.

### 3.1. *Monte Carlo sampling*

Monte Carlo (MC) methods are quite popular stochastic methods and the simplest to implement (see e.g. [12]). Basic concept of MC methods is a generation of set of outputs corresponding to a given set of inputs. It is further possible to apply sampling methods to estimate statistical nature of the outputs for a given variety on inputs (that can also be statistically sampled). The main advantage of MC methods is that it is sufficient to solve the deterministic model. Major limitation concerns the convergence rate, as the number of pairs of inputs and outputs needs to be relatively high in order to represent a statistically meaningful sample. Generation of this large samples may be computationally expensive.

Applying MC methods to inverse analysis is computationally even more costly, since inverse operator involves more computing time than a direct one, as it requires a large number of simulations. The use of MC in present study of inverse analysis based on indentation test is practically made possible by the employment of reduced basis model designed for fast simulation of the indentation test (following the procedure outlined in section 2). The applied MC method consists of the following steps:

(1) Simulation of the test is performed attributing to sought parameters some reference values in order to generate a 'pseudo-experimental' indentation curve.

(2) $M$ different curves, corresponding to presumed $M$ different experiments performed on the same specimen, are generated by affecting all measurable quantities in the indentation curve computed in (1) by the same perturbation coefficient ($nf$ for the force and $nd$ for corresponding penetration depth). Each point of the reference curve is modified such that, to the starting value of force a perturbation amounting to $nf\%$ of its previous value is added; the same for corresponding penetration depth using $nd\%$ of reference value. By extracting $M$ different 'noise' levels for $nf$ and $nd$ within some specified range (say ±5%) $M$ different indentation curves are created.

(3) *M* generated curves in (2) are sampled in order to obtain mean value of pseudo-experimental data, denoted by $\mathbf{d}_{mean}^{EXP}$ and corresponding standard deviation.

(4) A large number of indentation curves (say *S* of them) is generated using Gaussian distribution (or any other distribution) with mean value and standard deviation computed in (3), for each measurable quantity from pseudo-experimental curve. The number *S* should be large enough to obtain statistically meaningful set of indentation curves.

(5) Inverse analysis is performed using as an input each of *S* indentation curves generated in (4). Resulting material parameters are sampled to obtain their mean value and standard deviation.

It is worth noting that the computation burden connected to *S* inverse analysis (with *S* equal to 2000 or more) is significantly reduced by the use of reduced basis model specifically calibrated for the indentation test referred to in this study.

### *3.2. Extended Kalman filters*

Kalman filtering is developed in control engineering as a tool that provides a recursive solution to the linear optimal filtering problem for dynamic systems [13]. The developed algorithm can be used to produce estimates of unknown state variables with higher accuracy relying on series of measurements observed over time, instead of using a single measurement.

By applying Kalman filters to solve inverse problems it is possible to generate the covariance matrix of estimated parameters, starting from measured quantities endowed by quantified uncertainty. The original form of Kalman filters is intended to be used for linear systems, and hence cannot be directly applied for the problem at hand, as the inverse analysis based on indentation test is highly non-linear problem. Thus, extended versions of Kalman filters will be employed, in its modified form to take into account non-sequential experimental data. The governing equations used for the updating procedure are derived by linearization about current estimate.

Theoretical fundamentals and detailed derivations considering Kalman filtering methodology can be found in [14, 15]. In what follows a sequence of operations with proper updating equations adopted in this study is outlined.

Let $\mathbf{d}^{EXP}$ be a vector of experimentally measured quantities, here representing $R$ pairs of points (i.e. force and corresponding penetration depths) from the pseudo-experimental indentation curve, while $\mathbf{C}^{EXP}$ is an $R \times R$ covariance matrix referring to the uncertainty in the measurements. Iterative procedure used for the estimates in the parameters and corresponding covariance matrix is following steps listed below.

(1) A starting value is attributed to all $P$ sought parameters, collected in vector $\mathbf{p}^{GUESS}$, and a vector of residuals is computed using direct operator (i.e. numerical model for test simulation) according to

$$\mathbf{r}\left(\mathbf{p}^{GUESS}\right) = \mathbf{d}^{EXP} - \mathbf{d}^{COM}\left(\mathbf{p}^{GUESS}\right) \tag{3}$$

(2) Sensitivity matrix denoted by $\mathbf{L}$ is computed in the following form:

$$\mathbf{L} = \frac{\partial \mathbf{d}^{COM}\left(\mathbf{p}^{GUESS}\right)}{\partial \mathbf{p}^{GUESS}} \tag{4}$$

(3) A Kalman 'gain' matrix is computed by linearization about the current iterate point, preserving first-order members from the Tailor series. The equation for calculation of gain matrix then reads:

$$\mathbf{K} = \left[\mathbf{L}^T \cdot \left(\mathbf{C}^{EXP}\right)^{-1} \cdot \mathbf{L} + \mathbf{C}_P^{-1}\right]^{-1} \cdot \mathbf{L}^T \cdot \left(\mathbf{C}^{EXP}\right)^{-1} \quad (5)$$

resulting matrix **K** is of the order $P \times R$.

(4) Updating equation is used for the evaluation of 'new' parameter values using experimentally measured data, and computed gain matrix by the following expression:

$$\mathbf{p}^{NEW} = \mathbf{p}^{GUESS} + \mathbf{K}\left[\mathbf{d}^{EXP} - \mathbf{d}^{COM}\left(\mathbf{p}^{GUESS}\right)\right] \quad (6)$$

Expression in the square brackets represents already computed vector of residuals, so no further computational 'cost' is required for the calculation of (6)

(5) A $R \times R$ covariance matrix of estimated parameters is evaluated using the equation given by:

$$\mathbf{C}_P = \left[\mathbf{L}^T \cdot \left(\mathbf{C}_{EXP}\right)^{1} \cdot \mathbf{L}\right]^{-1} \quad (6)$$

Presented updating procedure results from a full derivation by attributing to the initial covariance matrix of the parameters a large value, causing its inverse to vanish. The presented scheme therefore can be used in the absence of the a priori knowledge of the uncertainty affecting initial 'guess' value of the parameters. In case when also the covariance matrix of initial parameters is specified at the beginning of identification, equation (7) is modified so that also this matrix is updated throughout the procedure (see [16] for details). In this study the most general procedure requiring least amount of data as inputs is tested.

Iterative procedure given by equations (3) to (7) is terminated by reaching convergence criteria, previously specified on minimal changes in parameters between two consecutive iterations. Result of this calculation is vector of resulting parameter $\mathbf{p}^{RES}$ endowed by appropriate covariance matrix $\mathbf{C}_P$, with elements on the main diagonal being a square roots of standard deviation of assessed parameters (i.e. $s_i = \sqrt{c_{ii}^P}$).

## 4. Comparative stochastic assessment of elasto-plastic material parameter by two different procedures

The computational exercises presented in what follows serve as a numerical validation of the proposed methodologies. Results obtained by two outlined procedures are comparatively presented. In both considered approaches (i.e. MC and KF) a reference is made to an indentation test by standard Rockwell conical indenter with $120^0$ opening angle and 200 μm tip radius. In a view of indenter's much larger rigidity with respect to the tested specimen it is modeled as a rigid surface. Such choice contributes to the reduction of computing time involved in test simulation.

The isotropy of the material to be tested suggests axially symmetrical model. Finite Element Model (FEM) adopted for the test simulation by the commercial code ABAQUS [17] is visualized in Figure 2. It exhibits the following features: unilateral contact between surface of a rigid indenter and modeled specimen surface with Coulomb friction coefficient equal to 0.05 (considered as known parameter here, alternatively can be identified together with other sought parameters, see e.g. [18]); large strain regime according to their implementation in ABAQUS [17], not discussed here; zero displacements on the boundaries common with the surrounding practically undisturbed zone (adopted here for simplicity instead of a more accurate alternative as statical condensation of the surrounding solid); 4365 four-node quadrilateral elements (hence 8878 degrees of freedom).

Material response of tested specimen is assumed to be elastic with plasticity following Drucker-Prager (DP) yield criterion, in a view of potential applications of tested procedures to metal-ceramic composite materials. The use of more general DP formulation instead of Hencky-Huber-Mises (HHM), which is included in former as a special case of it, is suited better for describing mechanical behavior of ceramic materials and composites with higher content of it (see [19], [20]). DP yield criterion therefore can be seen as a modification of HHM criterion in order to take into account pressure sensitivity. Yield surface according to DP criterion with linear hardening has the following form:

$$\Phi = \sqrt{J_2(\mathbf{s})} + \alpha I_1 - k - h\lambda \leq 0 \tag{8}$$

where $\mathbf{s}$ is deviatoric stress tensor and $J_2$ its second invariant, $I_1$ is the first invariant of stress tensor $\boldsymbol{\sigma}$, $\alpha$ is the internal friction angle, $k$ is the cohesion, while the last term represent contribution of linear strain hardening, with $h$ being a material parameter and $\lambda$ plastic strain-rate multiplier. By assuming associative plasticity (and therefore adopting that dilatancy angle is equal to internal friction angle) the number of constitutive parameters that govern plastic response following DP criterion is equal to 3. The internal friction angle $\alpha$ and cohesion $k$ depend on initial tensile and compressive yield limits, $\sigma_{t0}$ and $\sigma_{c0}$ as follows:

$$\alpha = \sqrt{3} \cdot \frac{\sigma_{c0} - \sigma_{t0}}{\sigma_{c0} + \sigma_{t0}}, \quad k = \frac{2}{\sqrt{3}} \cdot \frac{\sigma_{c0} \cdot \sigma_{t0}}{\sigma_{c0} + \sigma_{t0}} \tag{9}$$

For the material under investigation in this study constitutive parameters to assess by designed inverse analysis procedure, are Young's modulus ($E$), and additional three plastic parameters: $\alpha$, $\sigma_{c0}$ and $h$, while remaining parameters can be calculated using equation (9).

In order to computationally verify outlined procedures, a pseudo-experimental indentation curve is generated using designed FEM model attributing to the material parameters the following reference values: $E=80$GPa, $\sigma_{c0}=250$MPa, $\alpha=10^0$ and $h=1000$MPa. These values can be regarded as typical and representative for ceramic-metal composites ([19], [21]), and will be treated as "targets" for the further identification procedure. Maximal adopted indentation force is taken to be equal to 500N, which is a value large enough for this type of materials, to ensure global response of the tested specimen.

### 4.1. *Stochastic inverse analysis by Monte Carlo sampling*

A reference indentation curve is generated through test simulation attributing to parameters above listed values. In order to produce additional curves starting from a reference one, a procedure outlined in section 3.1 is used. Here, in particular, the following level of noise was used: $nf=\pm 2\%$, $nd=\pm 5\%$, with $M=15$ different random noise extractions resulting in 15 different curves. These curves are further sampled in order to obtain a mean curve and a corresponding standard deviation for each of 50 levels of force along loading part and 50 levels of force along unloading part.

In a subsequent step, $S=2500$ different indentation curves are generated using Gauss distribution based on previously calculated statistical values. For each of these curves an inverse analysis is solved cantered on minimization of a discrepancy function defined by function (1). The solution of this minimization problem is here achieved by means of traditional mathematical programming algorithm, an iterative "first-order" Trust Region Algorithm, in its version available in MATALB [22], and briefly outlined in section 2. Recursive test simulations are performed by a fast computational tool based on POD and RBF interpolation, summarized in section 2, and described in details in [7], [11]. Designed Reduced Basis (RB) model is "trained" once-for-all by a set of 2520 simulations confined by the following upper and lower bounds of sought parameters, defining a reasonable range for the type of materials referred to in this study, namely:

$$E = 55\text{GPa} \div 115\text{GPa} \tag{10a}$$

$$\sigma_{c0} = 150\text{MPa} \div 350\text{MPa} \tag{10b}$$

$$\alpha = 6^0 \div 20^0 \tag{10c}$$

$$h = 700\text{MPa} \div 2200\text{MPa} \tag{10d}$$

The use of RB model significantly reduced computational burden involved in solving 2500 inverse problems, as it provides results to the test simulation in a real-time, with controllable error that is kept on the level as of FEM calculation. Resulting parameters from these inverse analyses are sampled in order to obtain mean value and corresponding standard deviation and quantitative values of these are given in Table 1.

By observing results in the table in may be noticed that the mean value of all the assessed parameters is rather close to the target one with the largest error less than 1%, regardless, rather significant "noise" level used as perturbation of the reference curve. This corroborates the fact that inverse problem is well-posed, and the stochastic solution of it provides results with limited influence of level of perturbation on the mean values of estimates. On the other hand standard deviation of all estimated parameters are affected to a larger extent by the perturbed inputs. Still however, they may be regarded as reasonably small (i.e. all comparable to considered level of noise which was ±5%) proving stability of the proposed stochastic solution.

## 4.2. Extended Kalman filters for stochastic solution of inverse problem

Present identification problem is further solved by an updating procedure of extended Kalman filters, outlined in sub-section 3.2. Reference curve, used as input, is accompanied also by a covariance matrix $\mathbf{C}^{EXP}$ constructed using the same noise level as in the previous case. The matrix is calculated by attributing to all diagonal members values that correspond to standard deviation of particular level of force. Indentation curve here is divided into 50 points along loading part and 50 points along unloading part, therefore the resulting matrix $\mathbf{C}^{EXP}$ is 100×100 with diagonal elements equal to:

$$c_{ii}^{EXP} = S_{Fi}^2 \tag{11}$$

where $S_{Fi}$ is previously calculated standard deviation of $i^{th}$ level of force from the indentation curve. Stochastic solution of the inverse problem by KF leads to results that are summarized in Table 2.

Adopted updating procedure turned out to be quite efficient leading to mean values that are rather close to "target" one, with the largest error smaller than 1%, like in MC case. As pointed out in sub-section 3.2, used updating procedure is developed by linearization about current iterate, so the solution is reached iteratively. It is nonetheless fairly efficient as it involved only relatively small number of iterations. The identification procedure turned out to be insensitive to the choice of initial 'guess' value of parameters as it is solved couple of times starting from different parameter combinations, leading every time to the same solution. Such circumstance evidences well posedness of the identification problem. Two typical plots of parameter updating are visualized in Figure 3 and Figure 4. The same efficiency was noticed also in other initializations (i.e. number of iterations was less than 8), not presented here for brevity.

Values of standard deviation for all parameters are higher than those obtained by Monte Carlo sampling. This is due to the selected updating procedure in which initial covariance matrix for parameters is assumed as unknown, attributing to the $\mathbf{C}_P^0$ large value, causing its inverse to vanish. Adopting this scheme inevitably led to larger values of assessed parameter's standard deviation. An improvement of this outcome can be reached by using different Kalman filter updating scheme which takes into account initial value of parameter covariance matrix.

It is worth commenting that somewhat larger values of standard deviation obtained for friction angle and hardening coefficient are due to the limited influence of these parameters to the test outcome (i.e. indentation curve). Sensitivity analysis conducted in this study showed that by perturbing separately each of four parameters with same relative amount with respect to their reference values, more pronounced effect on the indentation curve is provoked with perturbation of Young's modulus and yield limit than with remaining two parameters. Such circumstance led to somewhat larger values of standard deviation for these parameters, with mean values still accurately identified corroborating the robustness of the identification procedure.

## 5. Conclusions and closing remarks

Indentation instruments are presently equipped with tools capable to provide the relationship between applied force and the consequent penetration depth. The output of such test is a so-called 'indentation curve' which can be used for characterization of diverse materials, see e.g. [1, 2, 3]. These characterization procedures provide constitutive parameter estimates only. In engineering practice it is advantageous to assess also the uncertainty affecting the final estimates as a consequence of various sources of errors embedded within identification procedure.

The goal of this study was to develop a procedure apt for characterization of metal-ceramic composites based on instrumented indentation test centered on stochastic solution of inverse problem. A reference here is made to a Drucker-Prager constitutive model, suitable to describe inelastic behavior of ceramic based composites, see e.g. [19], [20], with total number of parameters to assess equal to 4. Preliminary study conducted here relied on pseudo-experimental data, since the main purpose was to comparatively test different computational procedures as for their efficiency. Results presented in what precedes led to the following conclusions.

Numerical exercises presented in this paper showed that the identification of parameters entering into Drucker-Prager constitutive model based on indentation curve results in a well-posed inverse problem. Minimization of discrepancy function by traditional deterministic algorithms leads to exact values of 'target' parameters, proving its convexity and verifying that the data collected from the indentation curve are sufficient for this purpose.

In order to solve the inverse problem in stochastic framework, a Monte Carlo sampling method can be used with success, relying on deterministic solution to the inverse problem, but performed many times. The computational burden connected to this solution can be significantly reduced by the application of RB model of test simulation that is trained by 'heavy' computation done once-for-all, and which can be further routinely used to provide results in a real-time manner. With tremendous reduction of computing time achieved through such RB model it became possible to use MC sampling for this purpose, since the number of test simulations involved in it was more than 100.000. This stochastic solution proved the robustness of designed characterization procedure since all the standard deviations of assessed parameters were within the same level as perturbation 'noise' used on a reference curve.

The use of Extended Kalman filter developed by linearization about current iterate turns out to provide quite effective scheme for the stochastic solution of this characterization problem regardless significant nonlinearities present in it. Relying on sensitivity matrix (i.e. on calculation of first derivatives only) updating scheme ensures convergence with rather effective rate as the number of iterations required to reach global function minimizer is smaller than 10, and also proved to be unaffected by the selection of initial 'guess' value for the parameters. By involving significantly smaller number of test simulations with respect to MC solution, it is possible here to use Finite Element Model without a need to make recourse to RB model, so laborious 'training' phase required to calibrate RB model can be avoided.

Updating procedure based upon the assumption of the absence of reliable quantification of 'guess' parameter uncertainty, described by a large initial covariance, obviously leads to somewhat larger values of standard deviation of estimated parameters, when compared to values obtained by MC, here taken as reference.

Future promising developments in progress concern the use of conceptually similar updating procedure that takes into account initial parameter guess covariance matrix. Further on an alternative, so called 'unscented' Kalman filter [23], [24], that doesn't require Taylor's expansion to approximate the non-linear forward operator, and its potential application in the present context is also subjection of current research.

**ACKNOWLEDGMENT**
This study is related to a research project on new ceramic technologies and novel multifunctional ceramic devices and structures, within European Union's Seventh Framework Programme FP7/2007-2013/ under REA grant agreement number PITN-GA-2013-606878-CERMAT2, and the authors are gratefully acknowledging the financial support from it.

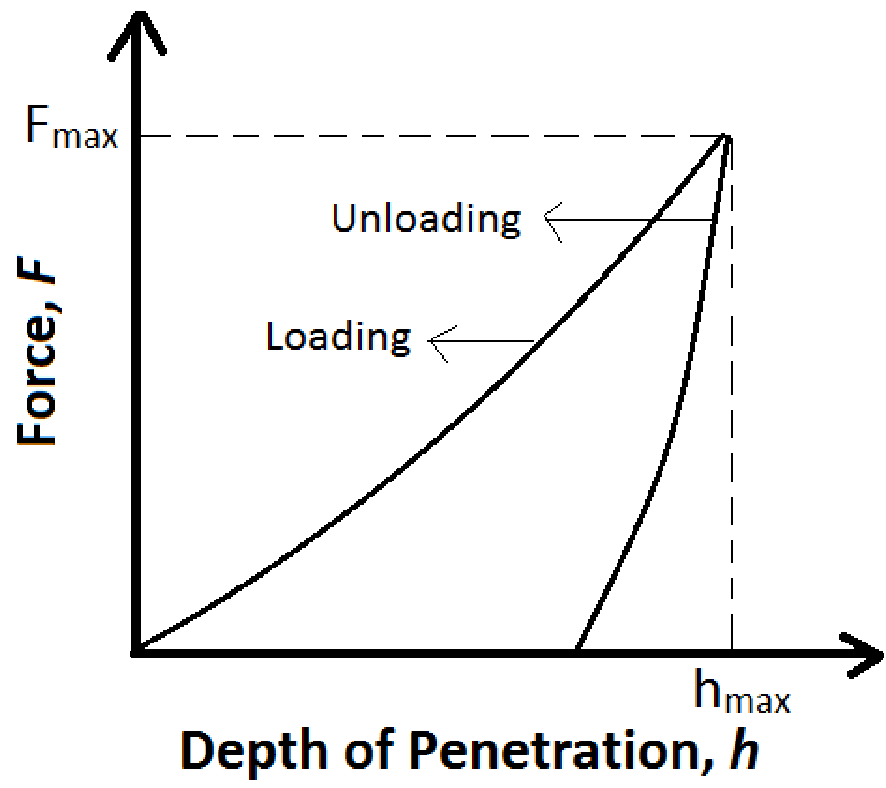

Figure 1. Indentation curve

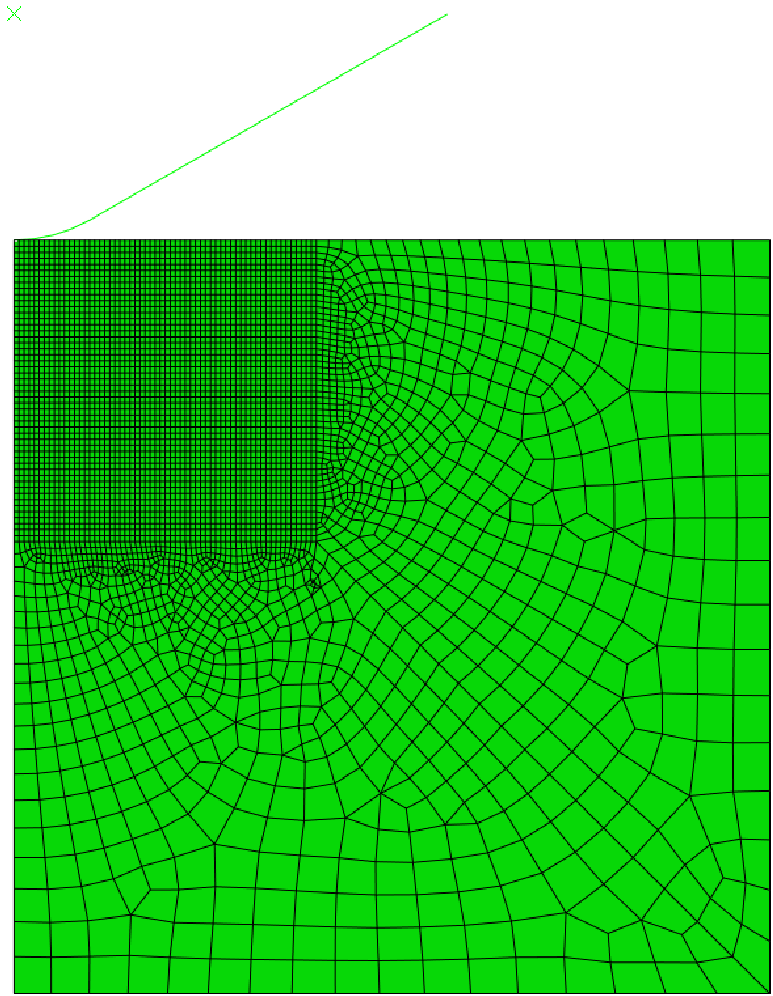

Figure 2. Axially symmetrical FE model for indentation test

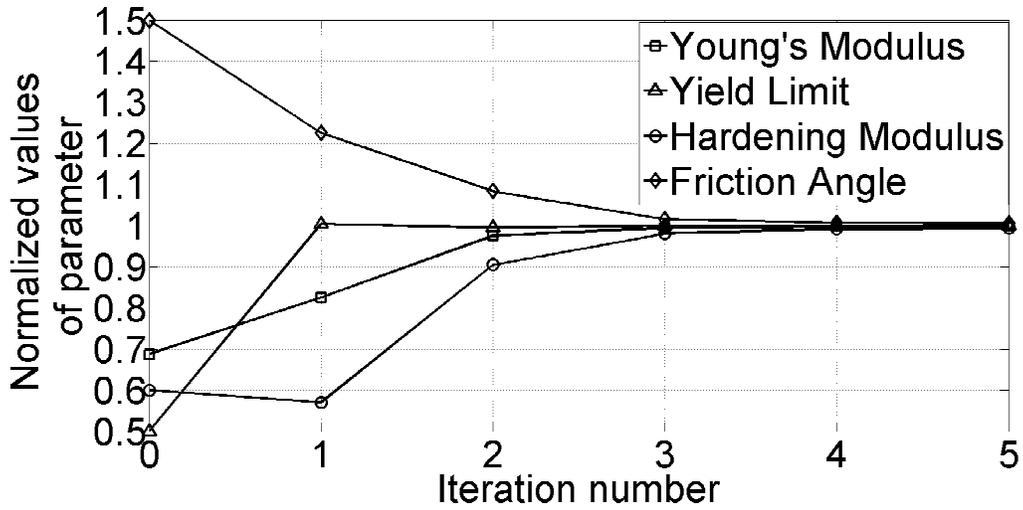

Figure 3. Normalized parameters updating values – initialization 1

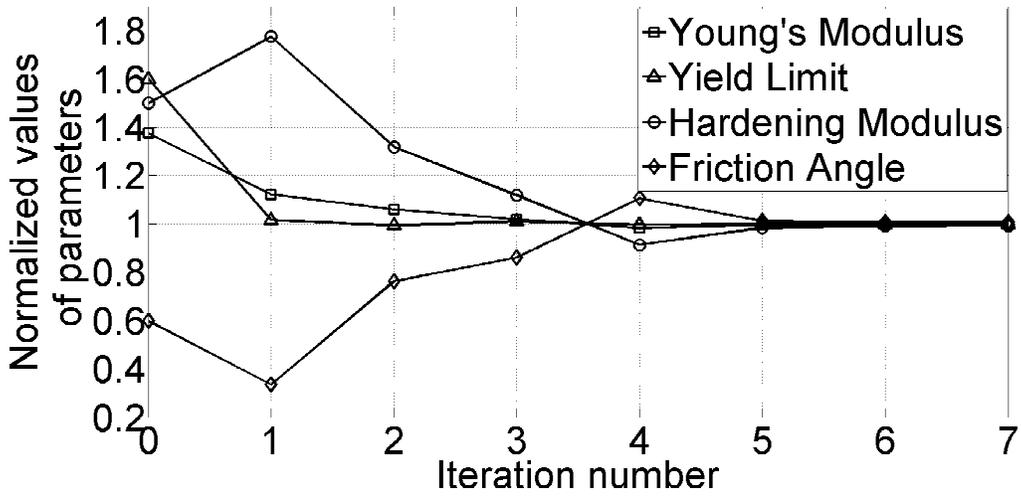

Figure 4. Normalized parameters updating values – initialization 2

Table 1. Results of inverse analysis by MC

| Parameter | Mean value | Standard deviation |
|---|---|---|
| $E$ | 79.83 GPa | ±2.39 GPa |
| $\sigma_{c0}$ | 249.85 MPa | ±4.67 MPa |
| $\alpha$ | $9.93^0$ | $±0.87^0$ |
| $h$ | 992.07 MPa | ±67.47 MPa |

Table 2. Results of inverse analysis by KF

| Parameter | Mean value | Standard deviation |
|---|---|---|
| $E$ | 79.81 GPa | ±7.58 GPa |
| $\sigma_{c0}$ | 250.23 MPa | ±7.27 MPa |
| $\alpha$ | $10.05^0$ | $±4.22^0$ |
| $h$ | 991.99 MPa | ±464.66 MPa |

**REFERENCES**

[1] Oliver, W.C. and Pharr, G. M.: An improved techniques for determining hardness elastic modulus using load and displacement sensing indentation experiments, *Journal of Materials Research*, Vol. 7, pp. 176-181, 1992.

[2] Bolzon, G., Buljak, V., Maier, G. and Miller, B.: Assessment of elastic-plastic material parameters comparatively by three procedures based on indentation test and inverse analysis, *Inverse Problems in Science and Engineering*, Vol. 19, No. 6, pp. 815-837, 2011.

[3] S., Kucharski and Z., Mroz: Identification of yield stress and plastic hardening parameters from a spherical indentation test, *International Journal of Mechanical Sciences*, Vol. 49, 1238–1250, 2007.

[4] Bolzon, G. and Buljak, V.: An indentation-based technique to determine in-depth residual stress profile induced by surface treatment of metal components, *Fatigue and Fracture of Engineering Materials and Structures*, Vol. 34, No. 2, pp. 97-107, 2011.

[5] Bhushan, B.: Handbook for micro/nano tribology, CRC Press, Boca Raton, 1999.

[6] ISO 6508-3:2005(E) Metallic Materials – Rockwell Hardness Test – Part 2: Verification and Calibration of Testing Machines (Scales A, B, C, D, E, F, G, H, K, N, T), ISO, Geneva, 2005.

[7] Buljak, V.: *Inverse analysis with model reduction: proper orthogonal decomposition in structural mechanics*, Springer Verlag, Berlin, 2012.

[8] Conn, A.R., Gould, N.I.M and Toint, P.L.: Trust-Region Methods, *Society for Industrial and Applied Mathematics (SIAM)*, Philadelphia, USA, 2000.

[9] Nocedal, J. and Wright, S.J.: *Numerical Optimization*, Springer Verlag, New York, 2000.

[10] Ostrowski, Z., Bialecki, R.A. and Kassab A.J.: Solving inverse heat conduction problems using trained POD-RBF network, *Inverse Problems in Science and Engineering*, Vol.16, No.1, pp. 705-714, 2008.

[11] Buljak, V. and Maier, G.: Proper Orthogonal Decomposition and Radial Basis Functions in material characterization based on instrumented indentation, *Engineering Structures*, Vol. 33, No. 2, pp. 492-501, 2011.

[12] Bruhns, O. and Anding, D.: On the simultaneous estimation of model parameters used in constitutive laws for inelastic material behaviour, *International Journal of Plasticity*, Vol. 15, pp. 1311-1340, 1999.

[13] Kalman, R.E. and Bucy, R.S.: New results in linear filtering and prediction theory, *ASME Journal of Basic Engineering*, Vol. 83, pp.95 – 108, 1961.

[14] Bittanti, S., Maier, G. and Nappi, A.: Inverse problems in structural elasto-plasticity: a Kalman filter approach, in: Sawczuk, A. and Bianchi, G., (Ed.): *Plasticity today, CISM book*, Elsevier Applied Science, pp. 311–329, London, 1984.

[15] Gibbs, B.P.: *Advanced Kalman Filtering, least-squares and modeling*, Wiley, New York, 2011.

[16] Fedele, R. and Maier, G.: Flat-jack tests and inverse analysis for the identification of stress states and elastic properties in concrete dams, *Meccanica*, Vol. 42, pp. 387-402, 2007.

[17] ABAQUS/Standard: Theory and user's manuals, release 6.5-1, HKS Inc., Pawtucket, 2005.

[18] Buljak, V., Cocchetti, G. and Maier, G.: Calibration of brittle fracture models by sharp indenters and inverse analysis, *International Journal of Fracture*, Vol. 184, pp. 123-136, 2013.

[19] Bolzon, G., Chiarullo, E.J., Egizabal, P. and Estournes, C.: Constitutive modelling and mechanical characterization of aluminium based metal matrix composites produced by spark plasma sintering, *Mechanics of Materials*, Vol. 3, 2010.

[20] Bocciarelli, M., Bolzon, G. and Maier, G.: A constitutive model of metal-ceramic functionally graded material behavior: formulation and parameter identification, *Computational Materials Science*, Vol. 43, pp. 16–26, 2008.

[21] Moon, R.J., Tilbrook, M., Hoffman, M. and Neubrand, A.: Al–Al2O3 composites with interpenetrating network structures: composite modulus estimation. *Journal of American Ceramic Society*, Vol. 88, pp. 666–674, 2005.

[22] The Math Works Inc., Matlab. User's guide and optimization toolbox, release 6.13, USA, 2002.

[23] Fedele, R., Maier, G. and Whelan, M.: Stochastic calibration of local constitutive models through measurements at the macroscale in heterogeneous media, *Computer Methods in Applied Mechanics and Engineering*, Vol. 195, pp. 4971-4990, 2006.

[24] Norgaard, M., Pulsen, N. and Ravn, O.: New developments in state estimation for nonlinear systems, *Automatica*, Vol. 36, pp. 1627–1638, 2000.